# On the negative capacitance in ferroelectric heterostructures

Yuchu Qin and Jiangyu Li[*]

Department of Materials Science and Engineering and

Guangdong Provincial Key Laboratory of Functional Oxide Materials and Devices

Southern University of Science and Technology

[*] Author to whom the correspondence should be addressed to; email: liju@sustech.edu.cn

**Abstract**

Negative capacitance can be used to overcome the lower limit of subthreshold swing (SS) in field effect transistors (FETs), enabling ultralow-power microelectronics, though the concept of ferroelectric negative capacitance remains contentious. In this work, we analyze the negative capacitance in ferroelectric/dielectric heterostructure rigorously using Landau-Denvonshire theory, identifying three (one) critical dielectric thicknesses for first (second) order ferroelectric phase transition upon which the stability of negative capacitance changes. A critical electric window is also identified, beyond which the ferroelectric negative capacitance cannot be maintained. Between the first and second critical thicknesses, meta-stable negative capacitance exists near zero polarization, yet it will be lost and cannot be recovered when the electric window is broken. Between the second and third critical thicknesses, stable negative capacitance always exists near zero polarization within the electric window regardless of initial polar state, resulting in hysteretic double P-E loop. Beyond the third (first) critical thickness of first (second) order phase transition, P-E loop becomes hysteresis free, though the spontaneous polarization can still be induced at sufficient large electric field. Singularities in the effective dielectric constant is also observed at the critical thickness or electric field. The analysis demonstrates that the negative capacitance of ferroelectric can be stabilized by linear dielectric within a critical electric window, and the negative capacitance can be either hysteresis free or hysteretic for first order ferroelectrics, while it is always hysteresis free for the second order ferroelectrics.

Negative capacitance has attracted a great deal of interests in recent years [1-7]. When a dielectric with negative capacitance is used as the gate in a field effect transistor (FET), subthreshold swing (SS) smaller than 60mV/dec [8], the fundamental limit set by the Boltzmann distribution, becomes possible [9], enabling ultralow-power microelectronics. Ferroelectrics possesses multiple wells in energy landscape, and thus negative capacitance naturally occurs on the top of their energy barriers [10], though it is thermodynamically unstable. Therefore, much efforts have been focused on stabilizing the negative capacitance of ferroelectrics by linear dielectrics [11,12], which was initially proposed by Salahuddin and Datta in 2008 [9]. Great progress has been made since then. For example, transient negative capacitances have been reported when ferroelectric energy barrier is crossed during polarization switching [13,14], while stable negative capacitance has been observed locally around the polar vortex core in a ferroelectric/dielectric superlattice [15]. Recently, suppressed polarization has also been observed in a ferroelectric field effect transistor (FeFET) with SS smaller than 60mV/dec [16], directly connecting negative capacitance of ferroelectrics at the material level to the device performance in FeFETs.

Despite these progresses, the idea of ferroelectric negative capacitance remains contentious, and alternative mechanisms exist for the apparent negative capacitance characteristics observed in experiments [17-20]. For example, it was suggested that in a transient measurement, positive capacitance of ferroelectrics [21] or reverse domain nucleation and propagation [22] could also explain the increased charge with decreased voltage, which was often attributed to transient negative capacitance. Sub-60 mV/dec SS can also arise in FeFETs from domain switching instead of negative capacitance, rending hysteresis in I-V curves that were widely observed in experiments [23,24] yet thought to be unexpected from ferroelectric negative capacitance. The original proposal of Salahuddin and Datta [9] was based on circuit analysis, yet T. P. Ma et al suggested it misapplied the Landau-Devonshire theory [17], arguing that under an external electric field the ferroelectric polarization is not equal to the free charge in FeFETs electrodes. Therefore, it is necessary to carefully reexamine the energetics of ferroelectric heterostructure at the material level, especially under external electric field, to see if and how ferroelectric negative capacitance can be stabilized, and to understand its implications.

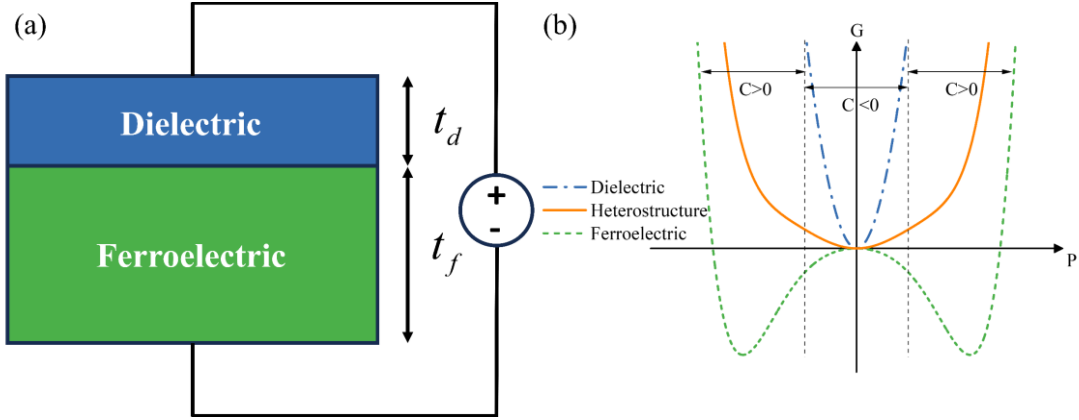

**Figure 1** Ferroelectric/dielectric heterostructure; (a) the schematic of structure; (b) the corresponding energy landscape.

We consider a ferroelectric/dielectric heterostructure as schematically shown in Fig. 1a, wherein the normalized thickness of ferroelectric and dielectric layers are $t_f$ and $t_d$, respectively, and both surfaces of the heterostructure are fully screened by the electrodes. The corresponding energy landscapes of the ferroelectric, dielectric, and heterostructure are shown in Fig. 1b, and it is seen that the heterostructure exhibits just one energy well with vanished polarization due to the presence of linear dielectric, illustrating the idea of stabilized ferroelectric negative capacitance in the heterostructure. Under an external electric field $E_0$, the free energy $G$ of the heterostructure per unit volume is given in term of polarization $P$ by [25],

$$G = \int [W(P) + \frac{\varepsilon_0}{2} E'^2 - E_0 P] dx \qquad (1)$$

where the first term of the integrand is the internal energy of the material system, the second term is the energy caused by the depolarization field $E'$ arising from discontinuity at the ferroelectric/dielectric interface, with $\varepsilon_0$ being vacuum permittivity, and the third term is the work done by the external electric field. Here the internal energy of the ferroelectric is given in term of polynomial of spontaneous polarization $P$ [26],

$$W(P) = \alpha_1 P^2 + \alpha_{11} P^4 + \alpha_{111} P^6 \tag{2}$$

where α are Landau coefficients, while the internal energy of dielectric is given in term of induced polarization,

$$W(P) = \frac{1}{2}\varepsilon_0 \varepsilon_d [\frac{P}{\varepsilon_0(\varepsilon_d - 1)}]^2 \tag{3}$$

with

$$P = \varepsilon_0(\varepsilon_d - 1)E_d \tag{4}$$

where $\varepsilon_d$ is relative dielectric constant and $E_d$ is the internal electric field of the dielectric. The internal field is contributed by both external field and depolarization field, which can be determined from Maxwell's equation

$$\nabla \cdot (\varepsilon_0 E + P) = 0 \tag{5}$$

with

$$t_d E_d + t_f E_f = E_0 \tag{6}$$

resulting in

$$E_d = \frac{t_f P}{\varepsilon_0(t_f \varepsilon_d + t_d)} + \frac{E_0}{(t_f \varepsilon_d + t_d)} \tag{7}$$

$$E_f = -\frac{t_d P}{\varepsilon_0(t_f \varepsilon_d + t_d)} + \frac{\varepsilon_d E_0}{(t_f \varepsilon_d + t_d)} \tag{8}$$

for dielectric and ferroelectric, respectively. As such, the free energy of the heterostructure per unit volume is derived as

$$G = t_f(\alpha_1 P^2 + \alpha_{11} P^4 + \alpha_{111} P^6) + \frac{t_d}{2\varepsilon_0(\varepsilon_d + \frac{t_d}{t_f})} P^2 - E_0 P \tag{9}$$

Note that in the derivation, we have simplified electric displacement *D* as polarization *P*, which is generally a good assumption for ferroelectric/dielectric heterostructure.

We first consider the ground state energetics of the heterostructure in the absence of external electric field. The equilibrium condition requires $\frac{\partial G}{\partial P} = 0$, such that

$$2t_f\alpha_1[1+\frac{t_d}{2\alpha_1\varepsilon_0(\varepsilon_d+\frac{t_d}{t_f})}]P+4t_f\alpha_{11}P^3+6t_f\alpha_{111}P^5=0 \tag{10}$$

which yield zero polarization with

$$P=0 \tag{11}$$

and nonzero polarization with

$$P^2=\frac{-\alpha_{11}\pm\sqrt{\alpha_{11}^2-3\alpha_{111}[\alpha_1+\frac{t_d}{2t_f\varepsilon_0(\varepsilon_d+\frac{t_d}{t_f})}]}}{3\alpha_{111}} \tag{12}$$

Furthermore, the stability requires that $\frac{\partial^2 G}{\partial P^2}>0$, such that

$$\alpha_1[t_f+\frac{t_d}{2\alpha_1\varepsilon_0(\varepsilon_d+\frac{t_d}{t_f})}]+6t_f\alpha_{11}P^2+15t_f\alpha_{111}P^4>0 \tag{13}$$

In the absence of dielectric layer, it is well known that zero polarization, and the corresponding negative capacitance associated with it, is unstable, because $\alpha_1$ is negative below Curie temperature [10], and thus ferroelectric polarization resides in one of the two energy wells given by Eq. (12). Nevertheless, the negative capacitance can be stabilized by the dielectric layer when the first critical thickness $\tau_1$ is reached,

$$\frac{t_d}{t_f}>-\frac{2\alpha_1\varepsilon_0\varepsilon_d}{2\alpha_1\varepsilon_0+1}=\tau_1 \tag{14}$$

beyond which there are three stable polarizations, though the energy at zero polarization is larger for first order ferroelectric phase transition where $\alpha_{11}<0$ [27,28], and thus the negative capacitance is metastable. Here we have considered the full polarization coupling between ferroelectric and dielectric, as emphasized by T. P. Ma et al [17]. Hoffmann et al derived a critical thickness without considering such coupling [2], though it is also a reasonable assumption since the denominator is very close to 1, which is also equivalent to the critical thickness identified by Salahuddin and Datta [9].

For first order ferroelectric phase transition, there also exists a previously unnoted second critical thickness $\tau_2$,

$$\frac{t_d}{t_f} > -\frac{2\varepsilon_0\varepsilon_d(\alpha_1 - \frac{\alpha_{11}^2}{4\alpha_{111}})}{2\varepsilon_0(\alpha_1 - \frac{\alpha_{11}^2}{4\alpha_{111}}) + 1} = \tau_2 \tag{15}$$

beyond which the energy at zero polarization becomes absolute minimum, and negative capacitance become stable. Eventually the spontaneous polarization will become unstable at the third critical thickness $\tau_3$,

$$\frac{t_d}{t_f} > -\frac{2\varepsilon_0\varepsilon_d(\alpha_1 - \frac{\alpha_{11}^2}{3\alpha_{111}})}{2\varepsilon_0(\alpha_1 - \frac{\alpha_{11}^2}{3\alpha_{111}}) + 1} = \tau_3 \tag{16}$$

beyond which ferroelectric polarization is completely suppressed.

Table 1 Coefficients of BaTiO$_3$ [29] and SrTiO$_3$ [30]

| Coefficients | Value | Units |
|---|---|---|
| $\alpha_1$ | $4.124(T-115)\times 10^5$ | J·m/C$^2$ |
| $\alpha_{11}$ | $-2.097\times 10^6$ | J·m$^5$/C$^4$ |
| $\alpha_{111}$ | $1.294\times 10^9$ | J·m$^9$/C$^6$ |
| $\varepsilon_d$ | 300 | / |

Considering now the heterostructure consisting of ferroelectric BaTiO$_3$ (BTO) and dielectric SrTiO$_3$ (STO) at the room temperature as an example, with their corresponding material parameters listed in Table 1, and the three critical thicknesses are evaluated as 0.208, 0.253 and 0.268, respectively. Representative energy profiles of the heterostructure are shown in Fig. 2a, while its effective dielectric constant (evaluated at zero electric field) as a function of dielectric thickness is shown in Fig. 2b. With the increased dielectric thickness, the evolution of the energy profile is evident when three consecutive critical thicknesses are crossed, going from two stable energy wells with spontaneous polarization to three (meta-)stable energy wells, and then to just one stable energy well with zero polarization, confirming the idea that ferroelectric negative capacitance can be stabilized by linear dielectric with appropriate thickness. For the effective dielectric constant, two branches are noted, one with zero polarization and the other with spontaneous polarization. They have singularities at first and third critical thicknesses, respectively, due to the change of stability between polar and nonpolar states, and anomalously high effective dielectric constant is observed at the first critical thickness for the heterostructure with zero polarization. The dielectric constant is also enhanced at the third critical thickness for the heterostructure with spontaneous polarization, though the magnitude is much smaller.

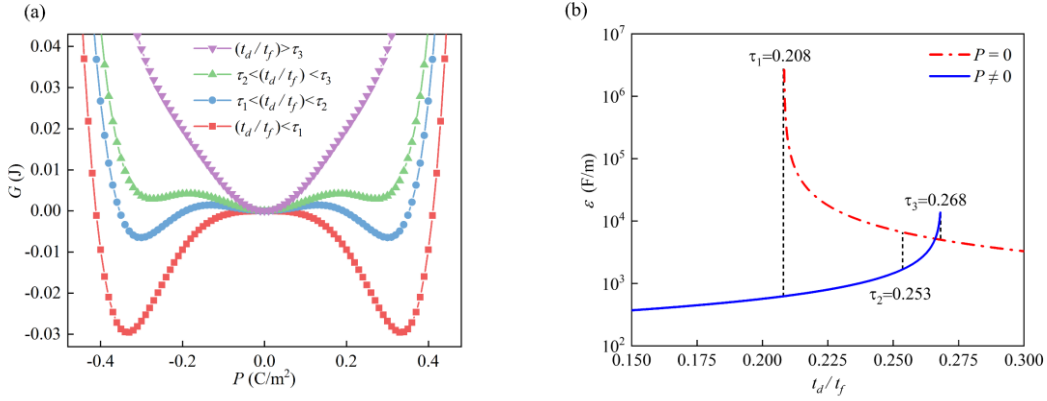

**Figure 2** The energy landscape (a) and effective dielectric constant (b) of the heterostructure with different dielectric thickness.

We then consider the evolution of negative capacitance under an external electric field, which is important since it could tip the subtle energetic balance between zero and spontaneous polarization. It is found that the stable negative capacitance can only exist within an electric field window, while initial polarization state (zero or nonzero) is also important as exhibited in Fig. 2b. In particular, for a dielectric between the first and third critical thicknesses, with initial polarization near zero having (meta-)stable negative capacitance, the electric field has to satisfy

$$E_0(P) = 2t_f \alpha_1 [1 + \frac{t_d}{2t_f \varepsilon_0 \alpha_1 (\varepsilon_d + \frac{t_d}{t_f})}] P + 4t_f \alpha_{11} P^3 + 6t_f \alpha_{111} P^5 \qquad (17)$$

and its derivative (or the second derivative of $G$) has four nonzero roots ranked in increasing order,

$$P(1) = -\sqrt{\sqrt{(\frac{\alpha_{11}}{5\alpha_{111}})^2 - \frac{\alpha_1}{15\alpha_{111}} - \frac{t_d}{30\alpha_{111}\varepsilon_0 t_f (\varepsilon_d + \frac{t_d}{t_f})}} - \frac{\alpha_{11}}{5\alpha_{111}}} \qquad (18)$$

$$P(2) = -\sqrt{-\sqrt{(\frac{\alpha_{11}}{5\alpha_{111}})^2 - \frac{\alpha_1}{15\alpha_{111}} - \frac{t_d}{30\alpha_{111}\varepsilon_0 t_f (\varepsilon_d + \frac{t_d}{t_f})}} - \frac{\alpha_{11}}{5\alpha_{111}}} \qquad (19)$$

$$P(3) = \sqrt{-\sqrt{(\frac{\alpha_{11}}{5\alpha_{111}})^2 - \frac{\alpha_1}{15\alpha_{111}} - \frac{t_d}{30\alpha_{111}\varepsilon_0 t_f (\varepsilon_d + \frac{t_d}{t_f})}} - \frac{\alpha_{11}}{5\alpha_{111}}} \quad (20)$$

$$P(4) = \sqrt{\sqrt{(\frac{\alpha_{11}}{5\alpha_{111}})^2 - \frac{\alpha_1}{15\alpha_{111}} - \frac{t_d}{30\alpha_{111}\varepsilon_0 t_f (\varepsilon_d + \frac{t_d}{t_f})}} - \frac{\alpha_{11}}{5\alpha_{111}}} \quad (21)$$

which correspond to extreme values of $E_0(P)$, upon which the stability of $G(P)$ switches. In fact $P(1)$ and $P(4)$ are polarizations at coercive field, while $P(2)$ and $P(3)$ are allowable polarization around zero with (meta-)stable negative capacitance. The window of electric field for stable negative capacitance are thus determined as

$$E_0(P(2)) \leq E_0 \leq E_0(P(3)) \quad (22)$$

For BTO/STO heterostructure, the correlation between initial polarization range having stable negative capacitance, and the window of electric field within which such stable negative capacitance can be maintained, are shown in Fig. 3 with respect to the dielectric thickness. Two branches of curves are noted again depending on the initial polarization state. With increased dielectric beyond the first critical thickness, it is evident that both polarization magnitude and electric window are widened, suggesting increased stability of negative capacitance.

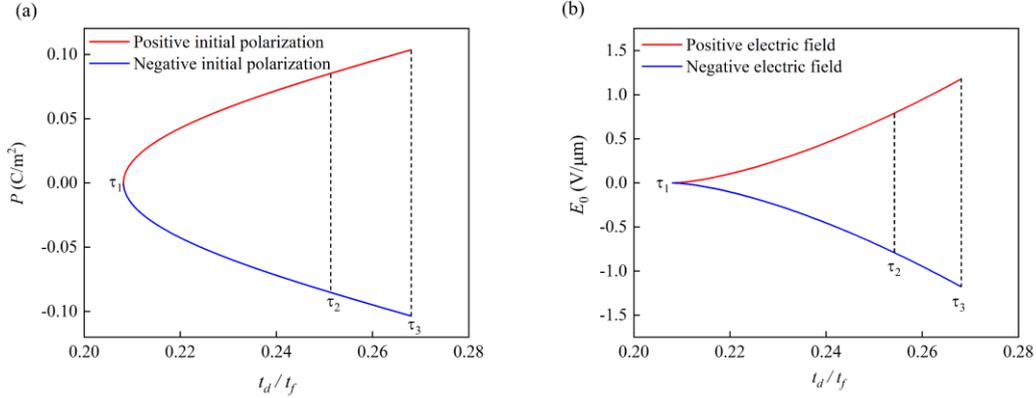

**Figure 3** Initial polarization range having stable negative capacitance (a) and the window of electric field within which such stable negative capacitance can be maintained (b).

To appreciate the critical role of the electric field, we evaluate the polarization of the heterostructure as function of external electric field for 4 representative thicknesses, as shown in Fig. 4. Between first and second critical thicknesses, both zero and spontaneous polarizations are (meta-)stable, and there are two branches of P-E loop as shown in Fig. 4a. One is hysteresis loop characteristic of typical ferroelectric with spontaneous polarization and switching, and the other is linear P-E curve close to zero polarization with sharp slope (and thus high effective dielectric constant) within a small electric window. When this electric window is broken, then polarization jumps to spontaneous one, and the stability of negative capacitance is lost. At the second critical thickness, the electric window coincides with the coercive field for polarization switching, as shown in Fig. 4b, and in both cases, the negative capacitance only exists within the small electric window and cannot be recovered if lost during electric cycling. Between the second and third critical thicknesses, however, the P-E becomes a double loop as shown in Fig. 4c, similar to antiferroelectric loop to certain extent, and it always goes through the negative capacitance region within the electric window during cycling, regardless of initial polar state. For example, in the forward sweep starting from the negative spontaneous polarization (and positive capacitance), it jumps to zero polarization and negative capacitance at zero electric field, and then follows the linear curve with negative capacitance when electric field is increased in the positive direction, before finally jumps to positive spontaneous polarization (and positive capacitance) beyond the electric field window. In the backward sweep it follows similar path. When the dielectric increases further beyond the third critical thickness, the hysteresis disappears in P-E loop (Fig. 4d), though spontaneous polarization can still be stabilized at sufficient large electric field, evaluated to be $\pm 4.94$ V/μm here from Eq. (22). This is reflected by the modest jump in the P-E loop beyond linear regime.

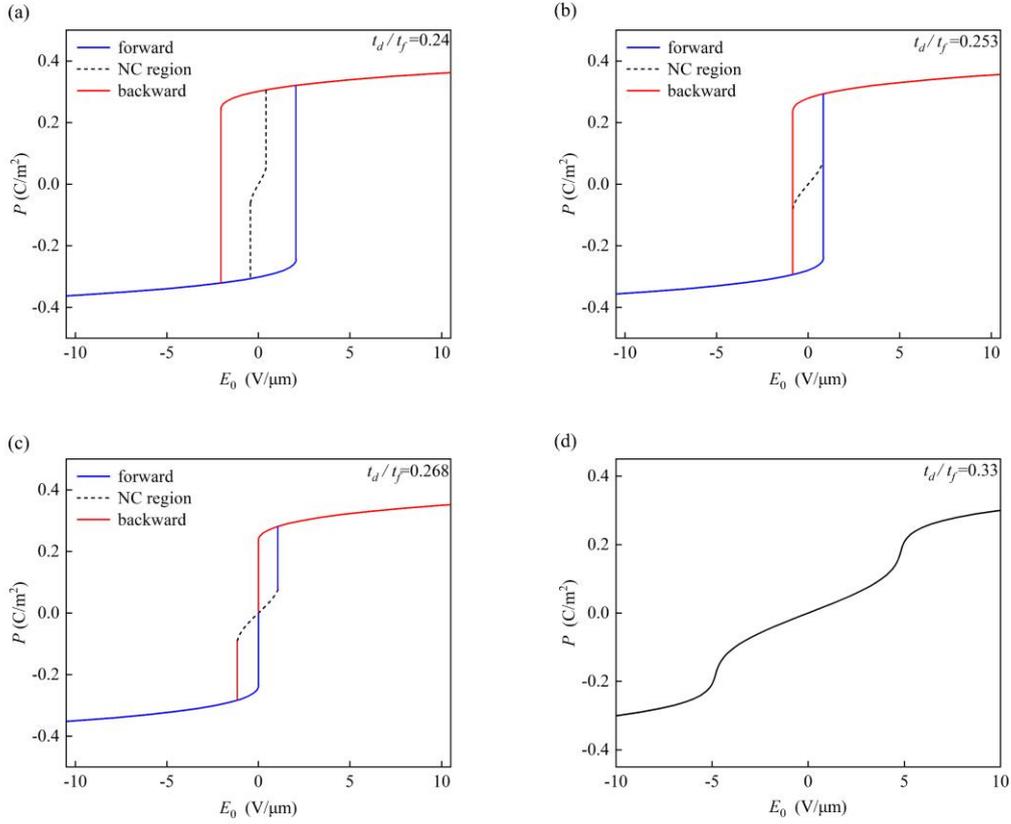

**Figure 4** P-E loops of the heterostructure for 4 representative dielectric thicknesses.

It is also revealing to examine the effective dielectric constant of the heterostructure versus the external electric field, as shown in Fig. 5. Between first and second critical thickness, classical butterfly loop is observed with sharp increase in dielectric constant at coercive field, but there is also a branch with large dielectric constant within a small electric window around zero field, corresponding to the metastable negative capacitance with near zero polarization (Fig. 5a). At the second critical thickness, this electric window coincides with the coercive field, and the dielectric constant increases sharply with increased electric field in either direction (Fig. 5b). In both cases, such high dielectric constant around zero field will be lost if the electric window is broken. Between second and third critical thicknesses, the negative capacitance around zero polarization becomes stable, and the dielectric-electric curve becomes more complex. Starting with a modest dielectric constant under negative electric field, it increases gradually when the field magnitude is reduced, followed by a rapid increase at zero field signaling switching of stability between zero and nonzero polarization. It then drops sharply upon breaking the electric window, beyond which the negative capacitance is lost. Again, the stable negative capacitance always exists within the electric window, regardless of the initial polar state. Beyond the third critical thickness, hysteretic butterfly loop disappears, and two symmetric double peaks emerge corresponding to field stabilized spontaneous polarization.

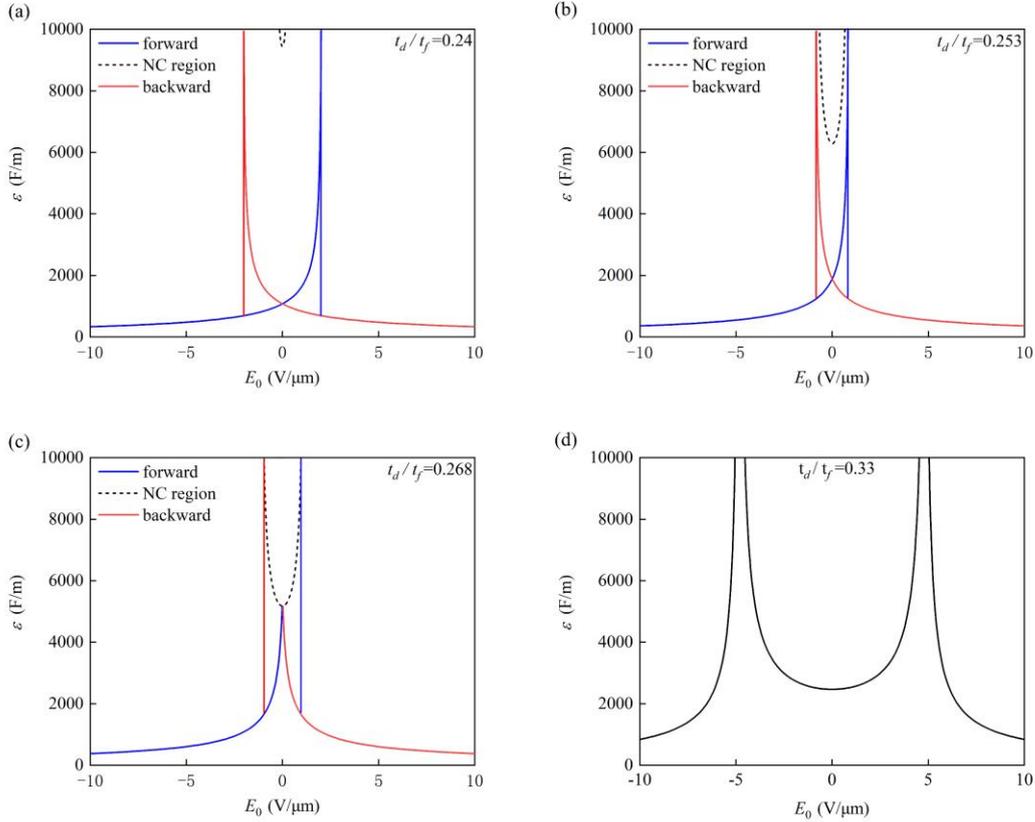

**Figure 5** The effective dielectric constant of the heterostructure versus external electric field for 4 representative dielectric thicknesses.

Finally, we note that for second order ferroelectric phase transition where $\alpha_{11}>0$ [27,28], there exists only one critical thickness, Eq. (14), beyond which the negative capacitance is stable within the electric field window specified by

$$E_0(P(1)) \leq E_0 \leq E_0(P(4)) \qquad (23)$$

P-E loop as well as dielectric curve are similar to Figures 4d and 5d, respectively. The behavior of negative capacitance in a second order ferroelectric thus is much simpler, and it is probably preferred in device applications.

**Acknowledgement**

We acknowledge the support of National Natural Science Foundation of China (92066203), Guangdong Provincial Key Laboratory Program (2021B1212040001) from the Department of Science and Technology of Guangdong Province. Li also acknowledges Financial Support for Outstanding Talents Training Fund in Shenzhen.